# Crystal Structure and Elastic Properties of Hypothesized MAX Phase-like Compound $(Cr_2Hf)_2Al_3C_3$

Yuxiang Mo*, Sitaram Aryal, Paul Rulis, and Wai-Yim Ching

Department of Physics and Astronomy, University of Missouri-Kansas City, Kansas City, Missouri 64110, USA

**Abstract**

The term "MAX phase" refers to a very interesting and important class of layered ternary transition-metal carbides and nitrides with a novel combination of both metal and ceramic-like properties that have made these materials highly regarded candidates for numerous technological and engineering applications. Using $(Cr_2Hf)_2Al_3C_3$ as an example, we demonstrate the possibility of incorporating more types of elements into a MAX phase while maintaining the crystallinity, instead of creating solid-solution phases. The crystal structure and elastic properties of MAX-like $(Cr_2Hf)_2Al_3C_3$ are studied using the Vienna *Ab initio* Simulation Package. Unlike MAX phases with a hexagonal symmetry ($P6_3/mmc$, #194), $(Cr_2Hf)_2Al_3C_3$ crystallizes in the monoclinic space group of $P2_1/m$ (#11) with lattice parameters of $a$ = 5.1739 Å, $b$ = 5.1974 Å, $c$ = 12.8019 Å; $\alpha = \beta = 90°$, $\gamma = 119.8509°$. Its structure is found to be energetically much more favorable with an energy (per formula unit) of -102.11 eV, significantly lower than those of the allotropic segregation (-100.05 eV) and solid-solution (-100.13 eV) phases. Calculations using a stress vs. strain approach and the VRH approximation for polycrystals also show that $(Cr_2Hf)_2Al_3C_3$ has outstanding elastic moduli.

Key words: MAX phases, crystals/crystallization, space group, elastic constants

* Corresponding author (yuxiangm@gmail.com);

## I. Introduction

The material under study, $(Cr_2Hf)_2Al_3C_3$, is derived from $Cr_2AlC$ which belongs to a large family of novel compounds: layered ternary transition-metal carbides and nitrides. The early discoveries[1–7] of these compounds (in powder form) date back to the 1960s. Three decades later, using a reactive hot-pressing method, Barsoum and El-Raghy[8] successfully fabricated single-phase $Ti_3SiC_2$ in polycrystalline bulk form, which enabled the observation of its unusual mechanical and transport properties. Such a development drew new attention to these carbides and nitrides, which have been extensively studied ever since[9–15]. The chemical compositions for most of these compounds can be summarized by a general formula: $M_{n+1}AX_n$ [M: an early transition-metal element, A: an A group (III, IV, V, or VI) element, X: carbon or nitrogen, and $n$ = 1 to 6], which is the reason why these compounds are also known as "MAX phases". In a MAX phase, hexagonal near close-packed M, A, and X atoms form nanolaminated layered structures. Figure 1(a) shows the unit cell of $Cr_2AlC$, which is sometimes called a "(2 1 1)" phase because the stoichiometric ratio of the constituent elements is 2:1:1. MAX phases involve a combination of metallic, covalent, and ionic bonds[16] among the composing atoms. Very recently, we made available[17] quantitative bond order values for the various species of bonds in MAX phases using Mulliken analysis. This uncommon blend of bonding types in MAX phases has given these nanolaminated materials a very intricate and intriguing combination of both metal and ceramic-like properties. They are good conductors of heat and electricity. They are light-weight, stiff, and refractory, but also easily machinable. They can tolerate external damages and internal defects, as well as thermal shocks and high-temperature oxidation. Ongoing and prospective applications of this remarkable group of materials include cutting tools[18], saws, brazed tools, nozzles, bearings, rotating parts in disk drives[10], tools for die pressing, biocompatible materials[11], electrodes, rotating electrical contacts[13], resistors, capacitors, heat exchangers[10], heating elements, kiln furniture, porous exhaust gas filters for automobiles, vacuum tube coatings in solar hot water systems, jet-engine components[19], coating materials on gas/steam turbine blades, nuclear applications[20, 21], projectile proof armor, fuselage materials of spacecraft to block infrared[22] and shield the craft from micrometeoroids and orbital debris, etc.

In addition to phases of the conventional $M_{n+1}AX_n$ formula, MAX phases with alternative stoichiometry and stacking sequence[23–25] have also been discovered. Other new materials in relation to MAX phases include ternary perovskite borides[26] and nitrides[27] (space group *Pm*3*m*, prototype $CaTiO_3$), $Al_3BC_3$[28] (a metal borocarbide containing linear C-B-C units), orthorhombic $Mo_2BC$[29] (space group *Cmcm*), the newly-discovered MXene[30] (2D nanosheets created via exfoliation of MAX phase), and so on. It has also been of genuine interest to investigate both experimentally (examples[31–42]) and theoretically[43–48] how adding more types of elements into a MAX phase would alter (and allow the tuning of) its properties, largely because the novel MAX phase itself was originally derived from conventional ceramics (binary transition-metal carbides and nitrides) by adding A group elements into them. Using $(Cr_2Hf)_2Al_3C_3$ as an example, we demonstrate, starting from a crystallographic point of view, the possibility of incorporating more types of elements into a MAX phase while maintaining the crystallinity, instead of creating solid-solution phases. Such preservation of crystallinity could not only provide exotic mechanical properties (as shown later) but also preserve tribological advantages[49, 50] and high electrical and thermal conductivities (due to the ease of electron and phonon scattering through periodic lattices). Compared with the solid-solution, a crystalline phase with a lower coefficient of friction can have a critical advantage in friction-reduction, self-lubrication, and wear-resistant applications. With a higher electrical conductivity, it can have better performance and less energy dissipation in electrical applications. And with a higher thermal conductivity, it can better protect itself (by efficient elimination of temperature gradients) against thermal shocks in high-temperature structural applications.

## II. Crystal Structure

In the present work, the combination of Cr and Hf was used to study novel changes of properties because these two elements have such a large contrast of atomic radii and numbers of valence electrons. In practice, many other elements could be adopted for the combination, forming a large series (detailed later) of potential new materials. Figure 1 delineates how the crystal structure of $(Cr_2Hf)_2Al_3C_3$ is developed from that of $Cr_2AlC$. The central idea is the replacement of Cr atoms at the center of each hexagon with another type of atoms (Hf, in the present case). This is illustrated schematically by the replacement of Fig. 1(d) with Fig. 1(e). Stoichiometry-wise, each hexagon in Fig. 1(e) uniquely includes 1 Hf atom at its center, while leaving 6 Cr atoms at the 6 vertices. But each of these Cr atoms is simultaneously shared by another 2 adjacent hexagons. So for each hexagon, there are $6\times\frac{1}{1+2}=2$ Cr atoms. The ratio for Cr and Hf in the crystal is 2:1, hence the chemical formula $(Cr_2Hf)_2Al_3C_3$.

To relax the crystal structure of $(Cr_2Hf)_2Al_3C_3$, we used the Vienna *Ab initio* Simulation Package (VASP)[51–53], with the implementation of the projector-augmented-wave (PAW) approach and the Perdew-Burke-Ernzerhof (PBE) exchange-correlation functional. An energy cutoff of 600 eV was set for the PAW-PBE potential and a Monkhorst Γ-centered 7×7×3 *k*-point mesh was used. The Methfessel-Paxton scheme was employed for the smearing of the Fermi surface. The electronic and ionic-force iterations converged to the level of $10^{-6}$ eV and $10^{-4}$ eV/Å, respectively. Detailed in Figure 2 is the VASP-relaxed crystal structure of $(Cr_2Hf)_2Al_3C_3$. Comparing it with the initial model in Figs. 1(f-h), the Hf atoms have deviated from the Cr planes and formed new sub-layers, mainly due to their larger atomic radii. Such extrusion of Hf atoms caused the shifting of Al atoms. An example is the Al atom ($x$=0.1598, $y$=0.4908, $z$=0.75) in layer "2" of Fig. 2, which was originally located at the $x$=0 cell boundary. The overall crystal lattice is also distorted by the atomic size difference. Constants for the new lattice are: $a$ = 5.1739 Å, $b$ = 5.1974 Å, $c$ = 12.8019 Å; $\alpha = \beta = 90°$, $\gamma = 119.8509°$.

It is also noticeable from the $\gamma$ angle in Fig. 2 that unlike the predecessor, MAX phase, $(Cr_2Hf)_2Al_3C_3$ does not belong to the hexagonal symmetry group anymore. Figure 3 is a plot of the crystal structures with extra colored lines that help illustrate the symmetry group for $(Cr_2Hf)_2Al_3C_3$. In Figure 3(a), only when the lattice is rotated by 180° (so that the blue pseudo-rhombus revolves to overlap with the magenta pseudo-rhombus) could the rotated lattice overlap with the original lattice after a $c/2$ translation in the $z$ direction. Any smaller rotation (to the yellow rhombus or green pseudo-rhombus) would not be able to achieve the overlap after translations, because such a rotation would move Hf atoms to the locations originally possessed by Cr atoms. So far it can be determined that the primitive cell has a 2-fold (360°/180°) screw axis with a translation of 1/2 of the $c$ lattice vector. In the Hermann-Mauguin notation, this is denoted as "$P2_1$". To find possible mirror planes and glide planes for $(Cr_2Hf)_2Al_3C_3$, a look first at $Cr_2AlC$ (the progenitor) would be very helpful. In Figure 3(b), there are two mirror planes: the vertical plane in green (crossing C, Al, and Cr atoms) and the horizontal plane in blue (crossing only Al atoms). The crystal structure stays unchanged after reflections of every atom according to either of the mirror planes. Here the unit cell of $Cr_2AlC$ is chosen to have the Al atom at the $z$=0 cell boundary, for the purpose of showing the symmetry associated with the horizontal mirror plane. Figure 3(c) shows the glide plane in purple. Reflections of every atom according to this plane result in a structure that needs a $c/2$ translation in the $z$ direction before it can overlap with the original structure. The two mirror planes and one glide plane shown in Figs. 3(b) and 3(c) are seen as "*mmc*" in $P6_3/mmc$, the Hermann-Mauguin notation for the space group of $Cr_2AlC$. From the unit cell of $(Cr_2Hf)_2Al_3C_3$ (which also has Al atoms at its $z$=0 cell boundary) in Fig. 3(d), it can be observed that the reflection invariance for the horizontal mirror plane still holds, but that for the vertical mirror plane does not exist anymore. Returning back to Fig. 3(a), the thin dotted line in green which crosses 4 C atoms indicates the hypothetical vertical mirror plane. It is apparent that Hf and Cr atoms are not mirror images to each other. Meanwhile, $a$ is not equal to $b$, so they would not be mirror images either even if they were the same type of atoms. A similar reasoning can easily tell that the hypothetical glide plane denoted by the thin dotted line in purple is not appropriate either. What remains in the case of $(Cr_2Hf)_2Al_3C_3$ is only one horizontal mirror plane. And this leads us finally to the complete Hermann-Mauguin notation: $P2_1/m$ (#11, in the monoclinic space group). With the relaxed structure of $(Cr_2Hf)_2Al_3C_3$, we used the PowderCell program[54] to simulate its X-ray (Cu, $K_\alpha$=1.540598 Å) diffraction pattern, which is plotted in Fig. 4. Such information provides a reference for future experimental confirmations of the phase. The electronic density of states is also calculated for $(Cr_2Hf)_2Al_3C_3$, using the first-principles orthogonalized linear combination of atomic orbitals (OLCAO) method[55]. The total energy was evaluated on a 7×7×3 $k$-point mesh in the irreducible portion of the Brillouin zone, and brought to convergence (0.0001 a.u. limit) in 81 iterations. Plotted in Fig. 5 is the electronic density of states for $(Cr_2Hf)_2Al_3C_3$. Like a MAX phase, this material is still metallic. Its total density of states (TDOS) at the Fermi level ($E_f$) is 18.65

states/(eV·cell). This fairly large value is primarily due to the relatively large number of atoms (24) in the unit cell. The major contributor to the TDOS at $E_f$ are the Cr-3*d* states [13.23 states/(eV·cell)], which can be attributed to the abundance of 3*d* electrons (5 per Cr atom) and the large number of Cr atoms (8 per unit cell).

### III. Total Energies and Elastic Properties

The energy convergence during the structural relaxation and the adequate stress under small strains (detailed in the next paragraph) suggest that the crystal structure of $(Cr_2Hf)_2Al_3C_3$ is at an energy minimum (structurally stable). This is the only condition always required for the ability of a material to exist. A structure does not necessarily have to possess the lowest energy among all the allotropic phases to be able to exist; otherwise there would not be any allotropes for any given chemical composition. In material systems that have relatively low energy barriers between different allotropic phases, less flexible preparation and heat treatment methods (and conditions), it is possible for the energy comparison to influence the purity and mass availability of a certain phase. To find out whether and to what extent $(Cr_2Hf)_2Al_3C_3$ is energetically favorable against its possible competing phases (that have the same chemical composition), additional VASP calculations of total energies were performed on six structures: a $(Cr_2Hf)_2Al_3C_3$ 1×1×3 supercell vs. a segregation model, a $(Cr_2Hf)_2Al_3C_3$ 3×3×1 supercell vs. a solid-solution model, $Cr_2AlC$ unit cell, and $Hf_2AlC$ unit cell. The structures of the first two sets of models are shown in Figs. 6 and 7, respectively. Listed in Table I are the numerical parameters and results of the calculations. Per formula unit, $(Cr_2Hf)_2Al_3C_3$ has an energy of -102.11 eV, significantly lower than those of the segregation (-100.05 eV), solid-solution (-100.13 eV), and $2(Cr_2AlC)+Hf_2AlC$ mixture (-101.04 eV). These energy differences of 1.07~2.06 eV per formula unit are quite large, equivalent to 103~199 $kJ{\cdot}mol^{-1}$. This indicates that the crystalline $(Cr_2Hf)_2Al_3C_3$ is much more preferable than the segregation, solid-solution, and $2(Cr_2AlC)+Hf_2AlC$ mixture, because large inputs of net energy are needed for Hf atoms in $(Cr_2Hf)_2Al_3C_3$ to form pure layers (segregation), break the order in Cr layers (forming solid-solution), or even for $(Cr_2Hf)_2Al_3C_3$ to completely degrade to $Cr_2AlC$ and $Hf_2AlC$. Meanwhile, $(Cr_2Hf)_2Al_3C_3$ is calculated to have a density of 7.58 $g/cm^3$, 4.26~6.76% higher than those of the segregation (7.25 $g/cm^3$), solid-solution (7.10 $g/cm^3$), and $2(Cr_2AlC)+Hf_2AlC$ mixture (7.27 $g/cm^3$). Therefore, $(Cr_2Hf)_2Al_3C_3$ is suggested to be resistant to not only temperature but also pressure-induced phase transitions.

To further compare the crystalline $(Cr_2Hf)_2Al_3C_3$ with the segregation and solid-solution in the mechanical perspective, the intrinsic elastic properties have also been calculated according to the strain-stress analysis scheme by Nielsen and Martin[56]. For each of these three phases and $Hf_2AlC$, a small compression (-1%) and expansion (+1%) were applied to each fully relaxed strain element. This compression/expansion rate was chosen as an optimal balance between linear elastic response and numerical accuracy. Keeping the volume and shape of these strained cells fixed, the atomic positions were fully relaxed again with the same computational configurations as those in Table I (segregation and solid-solution) and Sec. II [unit-cell $(Cr_2Hf)_2Al_3C_3$]. The resultant stress tensor ($\sigma_i$) ($i = xx, yy, zz, yz, zx, xy$) was used in combination with each corresponding strain ($\varepsilon_i$) to calculate the elastic tensor elements $C_{ij}$, according to the system of linear equations:

$$\sigma_{ij} = \sum_{ij} C_{ij}\varepsilon_j \tag{1}$$

Using the Voigt-Reuss-Hill approximation[57–59], the bulk mechanical parameters were then calculated from the elastic ($C_{ij}$) and compliance ($S_{ij}$) tensor elements. The Voigt approximation gives the upper limit for the parameters:

$$K_V = \frac{1}{9}(C_{11} + C_{22} + C_{33}) + \frac{2}{9}(C_{12} + C_{13} + C_{23}) \quad (2)$$

$$G_V = \frac{1}{15}(C_{11} + C_{22} + C_{33} - C_{12} - C_{13} - C_{23}) + \frac{1}{5}(C_{44} + C_{55} + C_{66}) \quad (3)$$

And the Reuss approximation gives the lower limit for the parameters:

$$K_R = \frac{1}{(S_{11} + S_{22} + S_{33}) + 2(S_{12} + S_{13} + S_{23})} \quad (4)$$

$$G_R = \frac{15}{4(S_{11} + S_{22} + S_{33}) - 4(S_{12} + S_{13} + S_{23}) + 3(S_{44} + S_{55} + S_{66})} \quad (5)$$

The Hill approximation as the average of the above two limits has been proven a good characterization of polycrystalline bulk properties of a variety of pristine[60–63] and defect materials[64–69]. The average of the two bounds is given by:

$$K = (K_V + K_R)/2, \qquad G = (G_V + G_R)/2,$$

$$E = 9KG/(3K + G), \text{ and } \eta = (3K - 2G)/2(3K + G) \quad (6)$$

Listed in Table II are the elastic coefficients, bulk moduli (K), shear moduli (G), Young's moduli (E), Poisson's ratio (η), and G/K ratio for the five phases. In terms of bulk moduli, Crystalline $(Cr_2Hf)_2Al_3C_3$ is elastically much stiffer than the allotropic segregation and solid-solution phases, as it has higher elastic moduli across the board: a bulk modulus of 181.5 GPa which is 19.6% and 21.9% larger than those of the segregation and solid-solution, a shear modulus of 125.2 GPa which is 19.5% and 40.7% larger than those of the segregation and solid-solution, and a Young's modulus of 305.3 GPa which is 19.4% and 37.1% larger than those of the segregation and solid-solution. To put these in the whole picture, in the drastic decline of elastic stiffness during the transit from $Cr_2AlC$ to $Hf_2AlC$, the segregation already has its elastic moduli almost comparable to those of $Hf_2AlC$. The solid-solution even has smaller shear and Young's moduli than those of $Hf_2AlC$. Yet stoichiometrically the segregation and solid-solution only have one third of the Cr atoms replaced with Hf atoms. While in sharp contrast, crystalline $(Cr_2Hf)_2Al_3C_3$ retains over 90% of the elastic stiffness of $Cr_2AlC$. As for the Poisson's ratio, most of these five phases fall in a fairly narrow range from 0.200 to 0.220 which is typical for MAX phases, while the solid-solution has a relatively higher (but still in-line) Poisson's ratio of 0.251. The Pugh ratio (G/K) for crystalline $(Cr_2Hf)_2Al_3C_3$ and the segregation phase are both 0.69, smaller than those of the $Cr_2AlC$ (0.73) and $Hf_2AlC$ (0.75). While the solid-solution phase has the smallest Pugh ratio of 0.60 which equals the $k_{crit}$ value[70] established for a fairly large group of MAX phases, indicating that the solid-solution phase is the least brittle among the five phases in the present study, but it is in the middle compared with other MAX phases.

There are various transition-metal elements available to play the roles of Cr and Hf in crystalline $(Cr_2Hf)_2Al_3C_3$. And the replacement of the central atom in each hexagon is not restricted to the M site only. The same pattern could be achieved at the A or X site. In addition, this type of phase does not have to be based on a (2 1 1) MAX phase. It could also come from (3 1 2) and (4 1 3) phases. Therefore, the general formula should be $(M1_2M2)_{n+1}(A1_2A2)(X1_2X2)_n$. Here "M1" and "M2" denote transition-metal elements. When $M1 \neq M2$, their stoichiometric ratio is 2:1, just like the case for Cr and Hf in $(Cr_2Hf)_2Al_3C_3$; When M1=M2, there is no replacement of M1 (in other words, the M site is pure). Exactly the same usage works as well for A1, A2, and X1, X2. And "*n*" represents the number of stacking layers. These proposed new crystals may inherit a significant portion of wonderful properties from MAX phases, but in the sense of crystallography they represent a new group of materials. We call them "D-MAX" phases because they can have "double" elements at one site. Although D-MAX phases currently await being experimentally synthesized and characterized, it is possible to surmise some general trends in their mechanical characteristics with respect to the structural and elemental variations. Enlarging the stacking number ($n$) of a D-MAX phase would make it behave more like a ceramic because of the increased stoichiometric content of the ceramic component. Replacing a transition-metal element (be it M1 or M2) with another transition-metal element of a comparable size but a larger number of valence electrons could strengthen the hybrid bonds and thereby stiffen the D-MAX phase. But replacing a transition-metal element (either M1 or M2) with another transition-metal element of the same number of valence electrons (isoelectronic) but a larger size would enlarge the bond lengths and weaken the D-MAX phase. Enlarging the difference in the atomic sizes of the two transition-metal elements (the larger atom being M2) would enlarge the energy differences between the D-MAX phase and its allotropes, provided the numbers of valence electrons are kept the same for each element. The same trends should also apply to the A site. And for the X site, the introduction of N would strengthen the original phase.

**IV. Summary and Conclusions**

We have studied the crystal structure and elastic properties of $(Cr_2Hf)_2Al_3C_3$ which is an example of a new type of MAX-like crystalline phases. It is found to be energetically much more preferable than competing segregation, solid-solution, and precursor structures. It also has outstanding elastic properties compared with the segregation and solid-solution phases, preserving over 90% of the elastic stiffness of $Cr_2AlC$ despite a 33.3% substitution of Cr by Hf atoms. According to the formula of D-MAX phases, and the possible MAX-phase elements[71] (M1, M2 = Sc, Ti, Zr, Hf, V, Nb, Ta, Cr, and Mo; A1, A2 = Cd, Al, Ga, In, Tl, Si, Ge, Sn, Pb, P, As, and S; X1, X2 = C and N; $n$ = 1 to 6) from the periodic table, a mathematical permutation and combination analysis suggests that there are 278640 potential D-MAX phases. Obviously, only a small portion of these could actually be synthesized. In the case of their progenitor, MAX phases, a similar analysis would predict that there can be 1296 phases, but so far the number of successfully[14] synthesized MAX phases is well below 100. Nevertheless, the present number of MAX phases has already fascinated many researchers and discoveries of new phases continue. We hope our work encourages others to not only experimentally synthesize members of the proposed D-MAX species via powder metallurgy and vapor deposition methods, but also theoretically predict which D-MAX phases, among the vast many, are more likely to be synthesized in pure samples by comparing the energy with that of the associated segregation, solid-solution, and precursor phases.

**Acknowledgments**

This work was supported by the US Department of Energy (DOE), National Energy Technology Laboratory under Grant No. DE-FE0005865. This research used the resources of the National Energy Research Scientific Computing Center supported by the Office of Science of DOE under Contract No. DE-AC03-76SF00098.

**Table I.** Computational configurations and results of $(Cr_2Hf)_2Al_3C_3$ supercells and corresponding models of segregation and solid-solution.

| | | 1×1×3 supercell | Segregation | 3×3×1 supercell | Solid-solution | $Cr_2AlC$ | $Hf_2AlC$ |
|---|---|---|---|---|---|---|---|
| | **Number of atoms** | 72 | 72 | 216 | 216 | 8 | 8 |
| | ***k*-point mesh** | 5×5×1 | 5×5×1 | 1×1×1 | 1×1×1 | 15×15×3 | 15×15×3 |
| **Convergence (eV/Å)** | **Electronic** | $10^{-6}$ | $10^{-6}$ | $10^{-6}$ | $10^{-6}$ | $10^{-6}$ | $10^{-6}$ |
| | **Ionic-force** | $10^{-4}$ | $10^{-4}$ | $10^{-4}$ | $10^{-4}$ | $10^{-4}$ | $10^{-4}$ |
| **Total energy (eV)** | **Per cell** | -612.79 | -600.29 | -1837.49 | -1802.26 | -65.32 | -71.43 |
| | **Per formula unit** | -102.13 | -100.05 | -102.08 | -100.13 | -101.04 | |
| **Lattice constants** | ***a*** **(Å)** | 5.17 | 5.21 | 15.56 | 15.75 | 2.85 | 3.27 |
| | ***b*** **(Å)** | 5.20 | 5.21 | 15.61 | 15.74 | 2.85 | 3.27 |
| | ***c*** **(Å)** | 38.41 | 39.88 | 12.79 | 13.38 | 12.69 | 14.39 |
| | ***α*** **(°)** | 90.00 | 90.00 | 90.00 | 90.07 | 90.00 | 90.00 |
| | ***β*** **(°)** | 90.00 | 90.00 | 90.00 | 90.15 | 90.00 | 90.00 |
| | ***γ*** **(°)** | 119.86 | 120.00 | 119.90 | 119.99 | 120.00 | 120.00 |
| | **Volume ($Å^3$)** | 895.55 | 937.50 | 2693.14 | 2872.91 | 89.27 | 133.26 |
| | **Density ($g/cm^3$)** | 7.59 | 7.25 | 7.57 | 7.10 | 7.27 | |

**Table II.** The elastic coefficients and intrinsic mechanical properties (in GPa) of $Cr_2AlC$, $(Cr_2Hf)_2Al_3C_3$ (unit cell), segregation, solid-solution, and $Hf_2AlC$.

| | $C_{11}$ | $C_{33}$ | $C_{44}$ | $C_{66}$ | $C_{12}$ | $C_{13}$ | $K$ | $G$ | $E$ | $\eta$ | $G/K$ |
|---|---|---|---|---|---|---|---|---|---|---|---|
| **$Cr_2AlC$[70]** | 364.5 | 356.1 | 139.8 | 140.0 | 84.4 | 107.4 | 187.0 | 136.1 | 328.6 | 0.207 | 0.73 |
| **$(Cr_2Hf)_2Al_3C_3$** | 355.2 | 333.6 | 120.1 | 136.6 | 80.9 | 107.4 | 181.5 | 125.2 | 305.3 | 0.220 | 0.69 |
| **Segregation** | 280.5 | 295.9 | 112.7 | 108.3 | 64.6 | 99.4 | 151.8 | 104.8 | 255.7 | 0.219 | 0.69 |
| **Solid-solution** | 286.3 | 237.6 | 86.6 | 102.4 | 77.3 | 94.2 | 148.9 | 89.0 | 222.7 | 0.251 | 0.60 |
| **$Hf_2AlC$** | 295.4 | 262.0 | 103.5 | 112.9 | 69.6 | 71.6 | 141.8 | 106.5 | 255.5 | 0.200 | 0.75 |

## Figure captions:

**FIG. 1.** The crystallographic evolution from $Cr_2AlC$ to $(Cr_2Hf)_2Al_3C_3$. (a) The unit cell of $Cr_2AlC$. (b) Cr-C layers separated from (a). (c) A 6×6 expansion of the middle layer in (b). (d) The same structure as that of (c), but with hexagons facilitating the observation of the hexagonal arrangements. (e) The structure from (c), with the central Cr atom in each hexagon replaced by Hf. (f) The same structure as that of (e), but with blue frames indicating the new unit cell. Derived in a similar approach from the top and bottom layers of (b), the left column of (g) shows the other two new layers, and the right column of (g) shows the new Al layers. (h) shows the preliminary unit cell of $(Cr_2Hf)_2Al_3C_3$, assembled with all the cell layers from (g) and (f).

**FIG. 2.** The relaxed crystal structure of $(Cr_2Hf)_2Al_3C_3$. Shown on the left is the unit cell of $(Cr_2Hf)_2Al_3C_3$ with purple numbers "1", "2", "3", "4", and "5" to its right, marking different layers. And shown on the right are the views of the five layers in the direction from +*z* to -*z*. Color strips contain the *z* parameters for each type of atoms. And numbers in red (upper) and green (lower) are the fractional *x* and *y* coordinates for individual atoms. Atoms without coordinates listed can be easily positioned according to symmetry.

**FIG. 3.** The space group symmetry of $(Cr_2Hf)_2Al_3C_3$. (a) The view (in the direction from +*z* to -*z*) of a 2×2×1 supercell of $(Cr_2Hf)_2Al_3C_3$. (b) The unit cell of $Cr_2AlC$, with two mirror planes in green and blue. (c) The unit cell of $Cr_2AlC$, with one glide plane in purple. (d) The unit cell of $(Cr_2Hf)_2Al_3C_3$, with one mirror plane in blue.

**FIG. 4.** Indexed X-ray diffraction pattern of $(Cr_2Hf)_2Al_3C_3$

**FIG. 5.** The electronic density of states for $(Cr_2Hf)_2Al_3C_3$. Total and atom-resolved DOS are shown on top, followed underneath by orbital-resolved DOS of each composing element.

**FIG. 6.** Crystal structures of the $(Cr_2Hf)_2Al_3C_3$ 1×1×3 supercell and segregation. The purpose of using a supercell is to include enough Hf atoms for the composition of an integer number of pure Hf layers in the segregation model.

**FIG. 7.** The construction of the solid-solution model based on the $(Cr_2Hf)_2Al_3C_3$ 3×3×1 supercell. The three-fold duplications in the *x* and *y* directions of a unit cell enable the shuffling of the Cr and Hf atoms.

**Figures:**

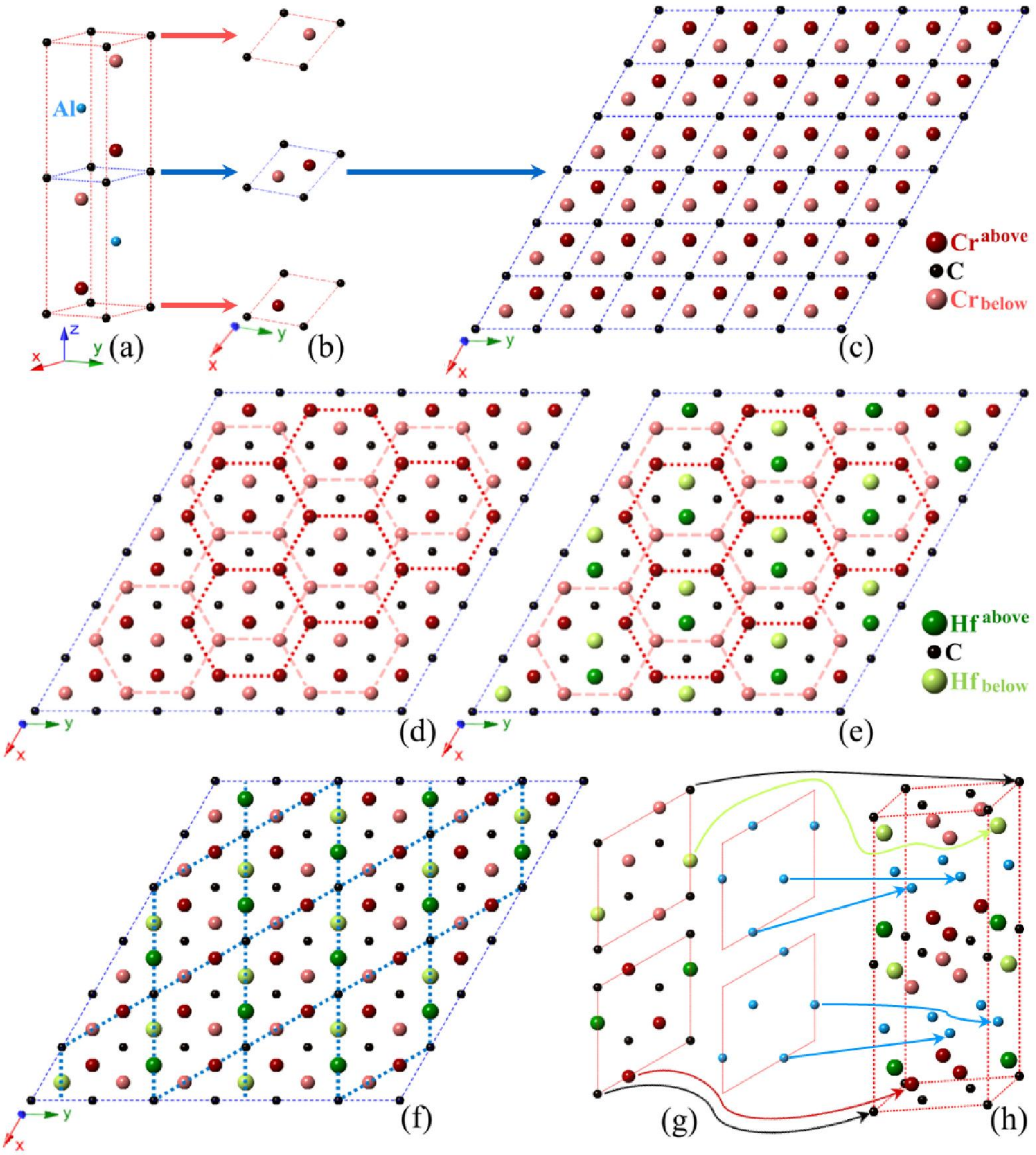

**FIG. 1.**

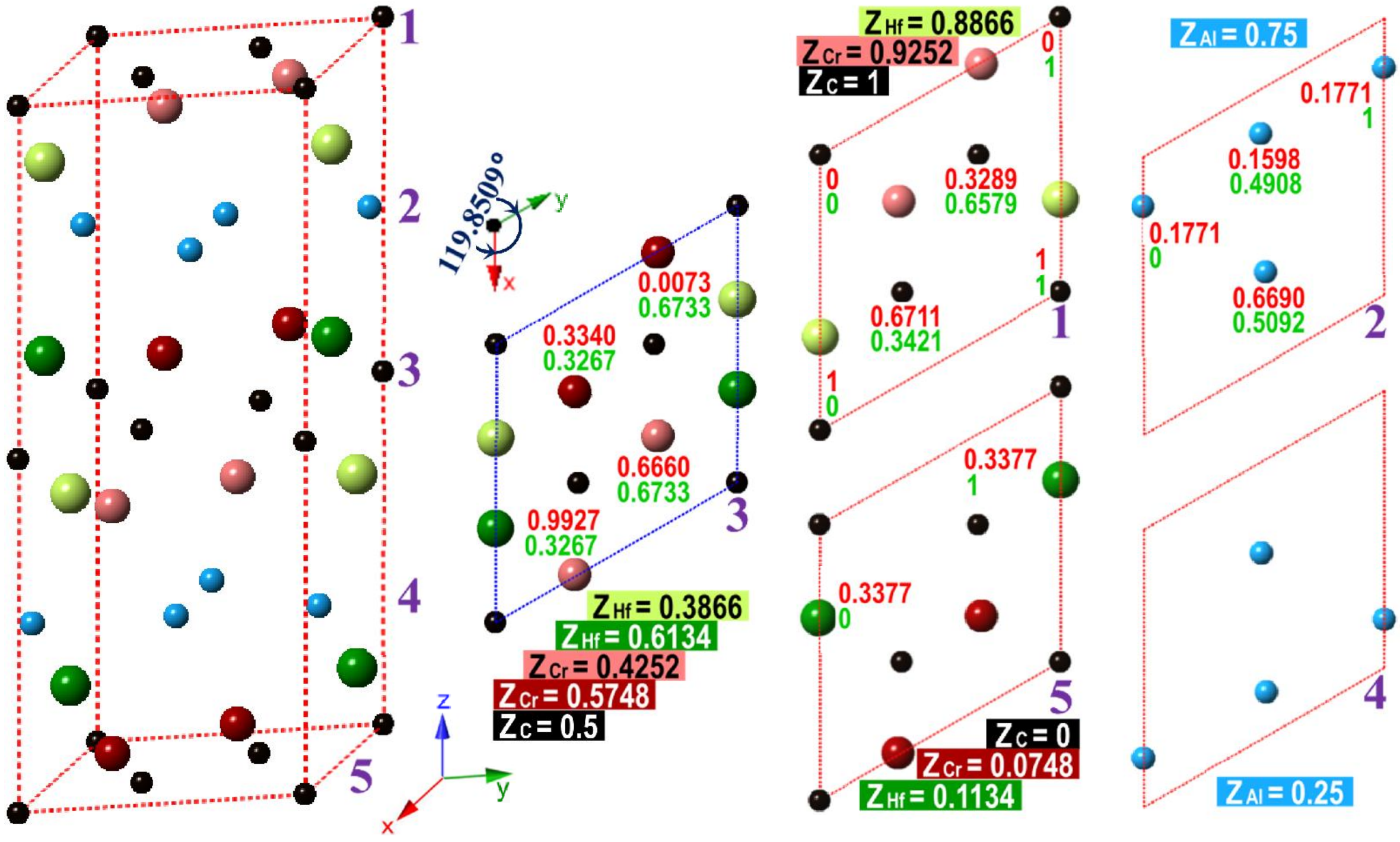

1
2
3
4
5
119.8509°
y
x
z
Z Hf = 0.3866
Z Hf = 0.6134
Z Cr = 0.4252
Z Cr = 0.5748
Z C = 0.5
0.0073
0.6733
0.3340
0.3267
0.6660
0.6733
0.9927
0.3267
Z Hf = 0.8866
Z Cr = 0.9252
Z C = 1
0.3289
0.6579
0.6711
0.3421
0.3377
Z C = 0
Z Cr = 0.0748
Z Hf = 0.1134
Z Al = 0.75
0.1771
0.1598
0.4908
0.6690
0.5092
Z Al = 0.25

**FIG. 2.**

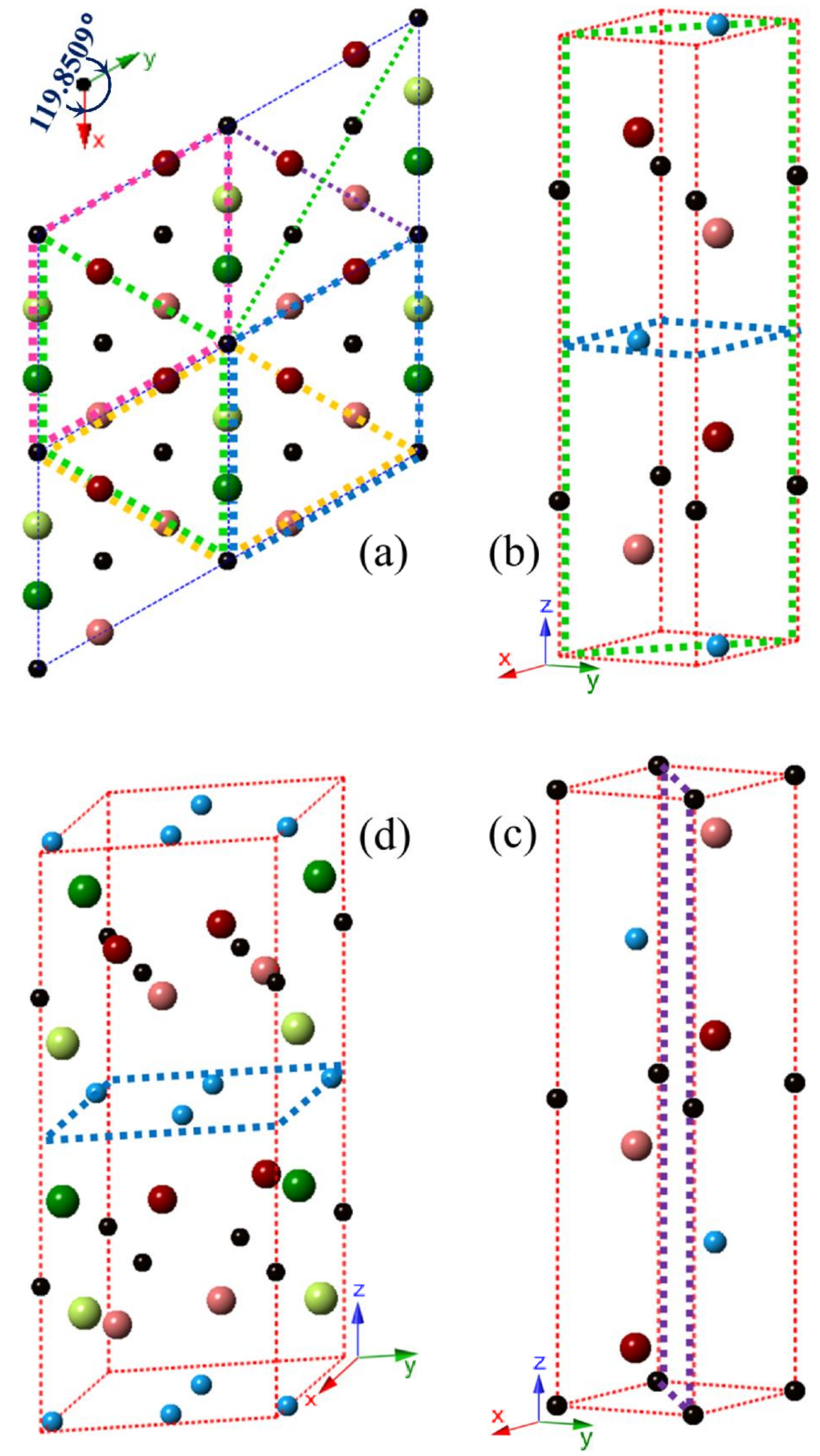

**FIG. 3.**

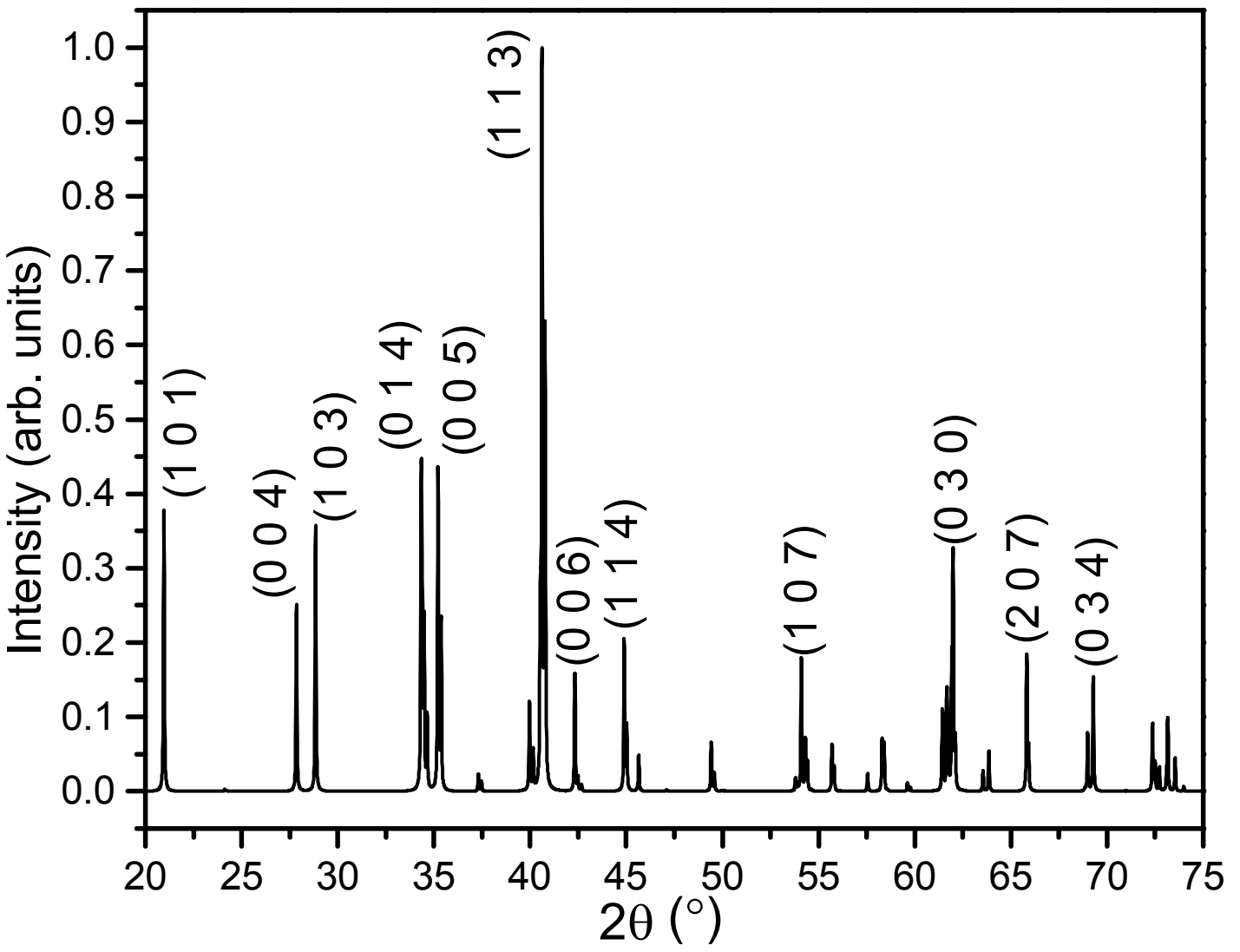

**FIG. 4.**

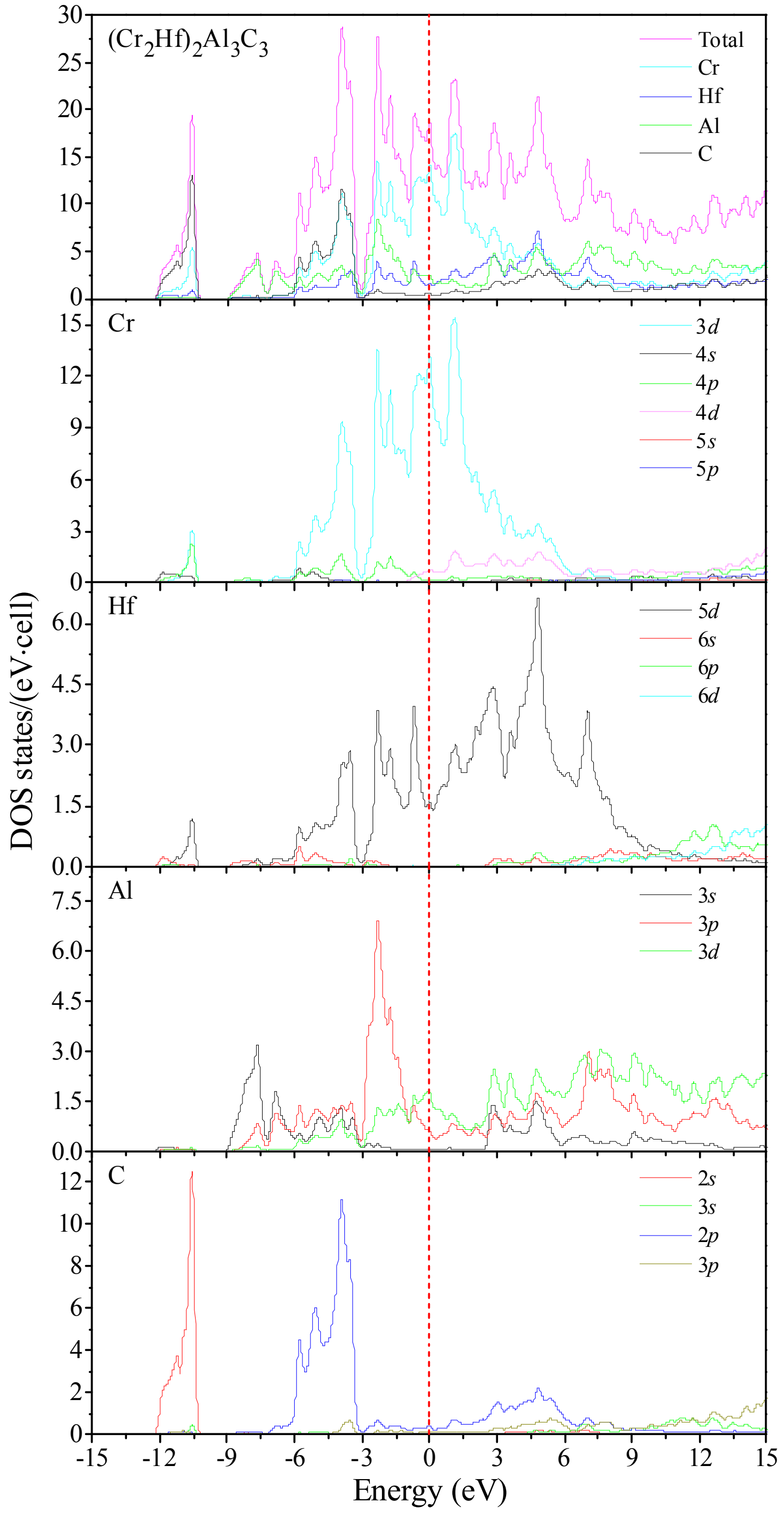

**FIG. 5.**

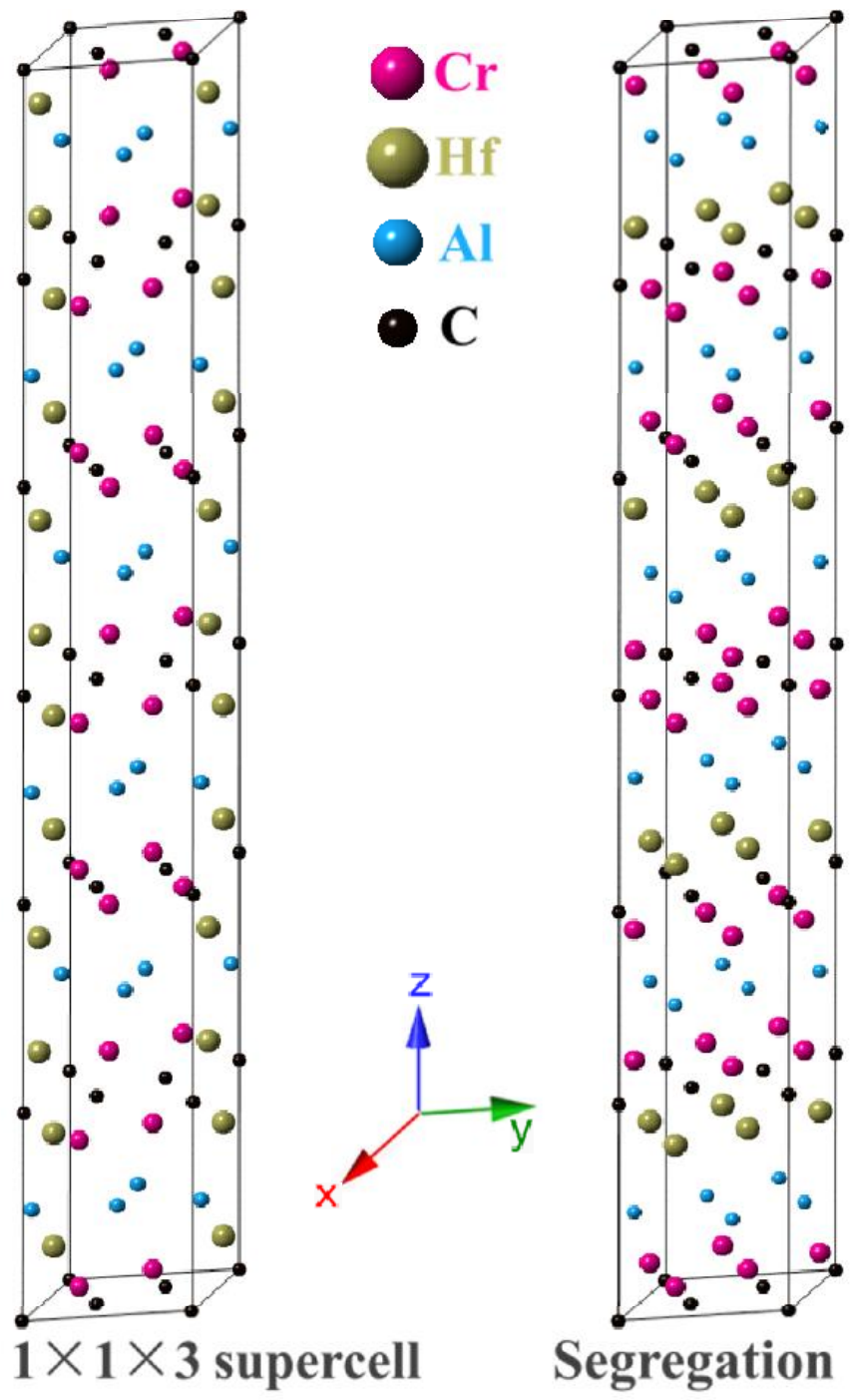

**FIG. 6.**

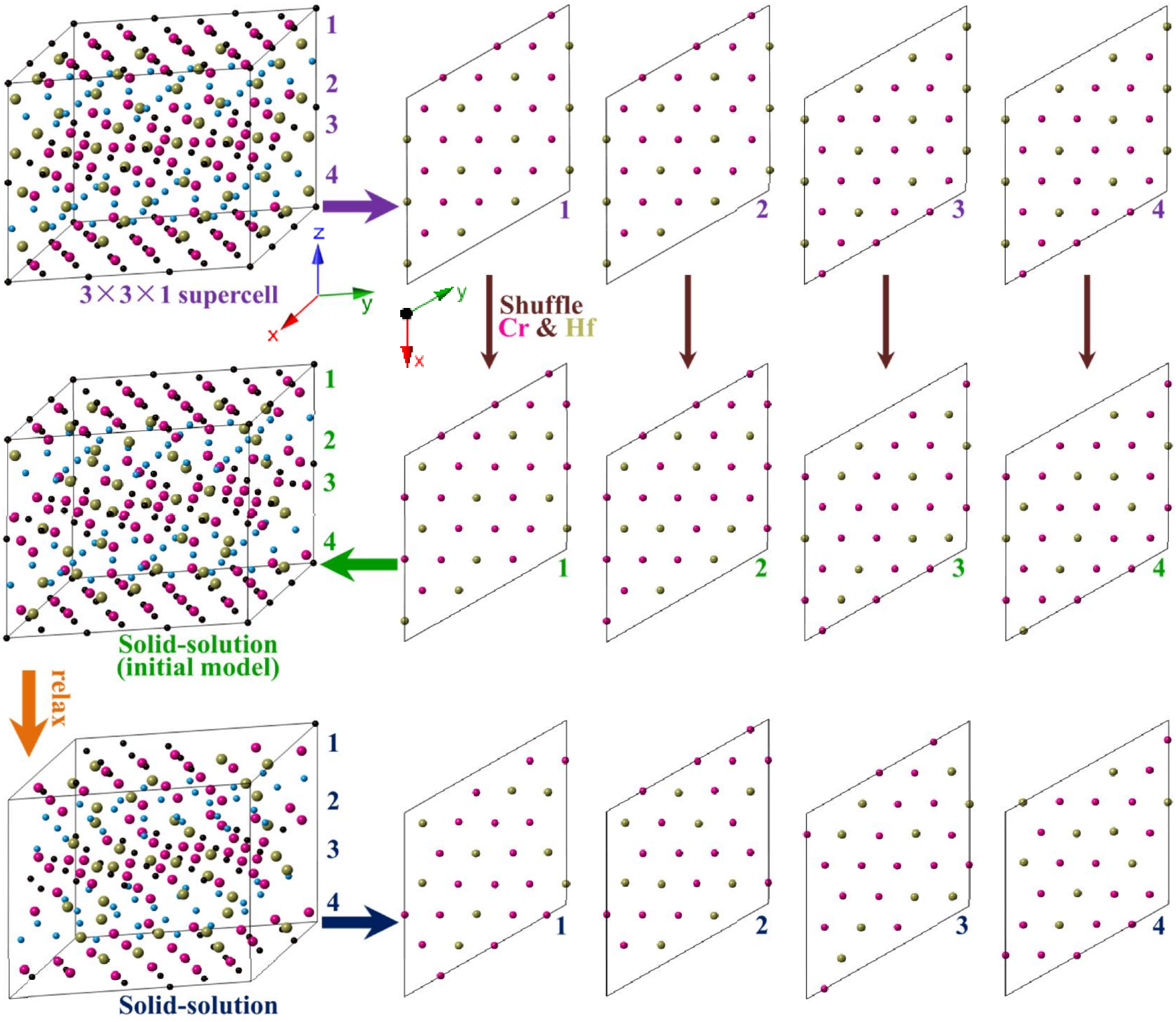

**FIG. 7.**